\documentclass[times,twocolumn]{aastex63}

\usepackage{CJK}
\usepackage{graphicx}
\usepackage{booktabs}
\usepackage{threeparttable}
\usepackage{amsmath, amssymb, bm}
\usepackage{mathrsfs}

\newcommand{\pmerge}{P_{\text{merge}}}

\newcommand{\hi}{\textrm{H}\textsc{I}}

\newcommand{\ha}{\ifmmode {\rm H}\alpha \else H$\alpha$\fi}
\newcommand{\hb}{\ifmmode {\rm H}\beta \else H$\beta$\fi}
\newcommand{\oiii}{[\textrm{O}\textsc{III}]}

\newcommand{\nii}{[\textrm{N}\textsc{II}]}

\newcommand{\sii}{[\textrm{S}\textsc{II}]}

\newcommand{\vasym}{v_{\text{asym}}}
\newcommand{\vasymbar}{\bar{v}_{\text{asym}}}
\newcommand{\verr}{\bar{v}_{\text{asym,err}}}
\newcommand{\logv}{\log \bar{v}_{\text{asym}}}

\newcommand{\dlogv}{\log \delta \bar{v}_{\text{asym}}}
\newcommand{\snr}{\text{SNR}(\re)}

\newcommand{\re}{R_{\text{e}}}

\newcommand{\logsfr}{\log \text{SFR}}
\newcommand{\mass}{M_\star}
\newcommand{\logm}{\log M_\star}

\newcommand{\ebv}{\text{E(B}-\text{V)}}

\newcommand{\qlocal}{Q_{\text{local}}}

\newcommand{\kpc}{\ h^{-1}\ \text{kpc}}
\newcommand{\mpc}{\ h^{-1}\ \text{Mpc}}
\newcommand{\kms}{\ \text{km}\ \text{s}^{-1}}
\newcommand{\pdis}{d_{\text{p}}}
\newcommand{\dv}{\Delta v}

\received{XX X, XXXX}
\revised{July 09, 2022}
\accepted{\today}
\submitjournal{ApJS}

\shorttitle{General Properties of Kinematic Asymmetry}
\shortauthors{Feng et al.}

\graphicspath{{./}{figures/}}

\begin{document}
\begin{CJK*}{UTF8}{gbsn}

\title{The Velocity Map Asymmetry of Ionized Gas in MaNGA \\ I. The Catalog and General Properties}

\correspondingauthor{Shuai Feng, Shi-Yin Shen}
\email{sfeng@hebtu.edu.cn, ssy@shao.ac.cn}

\author[0000-0002-9767-9237]{Shuai Feng (冯帅)}
\affiliation{College of Physics, Hebei Normal University, 20 South Erhuan Road, Shijiazhuang, 050024, China}
\affiliation{Hebei Key Laboratory of Photophysics Research and Application, 050024 Shijiazhuang, China}
\affiliation{Key Laboratory for Research in Galaxies and Cosmology, Shanghai Astronomical Observatory, Chinese Academy of Sciences, \\ 80 Nandan Road, Shanghai 200030, China}

\author[0000-0002-3073-5871]{Shi-Yin Shen (沈世银)}
\affiliation{Key Laboratory for Research in Galaxies and Cosmology, Shanghai Astronomical Observatory, Chinese Academy of Sciences, \\ 80 Nandan Road, Shanghai 200030, China}
\affiliation{Key Lab for Astrophysics, Shanghai 200234, China}

\author{Fang-Ting Yuan (袁方婷)}
\affiliation{Key Laboratory for Research in Galaxies and Cosmology, Shanghai Astronomical Observatory, Chinese Academy of Sciences, \\ 80 Nandan Road, Shanghai 200030, China}

\author{Y. Sophia Dai (戴昱)}
\affiliation{Chinese Academy of Sciences South America Center for Astronomy (CASSACA) / National Astronomical Observatories of China (NAOC), \\ 20A Datun Road, Beijing 100012, Peopleʼs Republic of China}

\author{Karen L. Masters}
\affiliation{Departments of Physics and Astronomy, Haverford College, 370 Lancaster Ave, Haverford, PA 19041}

\begin{abstract}

The SDSS-IV MaNGA survey has measured two-dimensional maps of emission line velocities for a statistically powerful sample of nearby galaxies. The asymmetric features of these kinematics maps reflect the non-rotational component of a galaxy's internal motion of ionized gas. In this study, we present a catalog of kinematic asymmetry measurement of $\ha$ velocity map of a sample of $5353$ MaNGA galaxies. Based on this catalog, we find that `special' galaxies (e.g. merging galaxies, barred galaxies, and AGN host galaxies) contain more galaxies with highly asymmetric velocity maps. However, we notice that more than half of galaxies with high kinematic asymmetry in our sample are quite `regular'. For those `regular' galaxies, kinematic asymmetry shows a significant anti-correlation with stellar mass at $\logm < 9.7$, while such a trend becomes very weak at $\logm>9.7$. Moreover, at a given stellar mass, the kinematic asymmetry shows weak correlations with photometric morphology, star formation rate, and environment, while it is independent of HI gas content. We also have quantified the observational effects in the kinematic asymmetry measurement. We find that both the signal-to-noise ratio of $\ha$ flux and disk inclination angle contribute to the measures of kinematic asymmetry, while the physical spatial resolution is an irrelevant factor inside the MaNGA redshift coverage.

\end{abstract}


\section{Introduction}\label{sec:intro}

Ionized gas, usually traced by emission lines in optical bands, is one of the key components of galaxies. Compared to the stellar components, ionized gas in disk galaxies is more sensitive to non-gravitational processes such as gas accretion, galactic wind, and AGN feedback, and then exhibits non-rotational motions such as radial flows and turbulence\citep{Veilleux2005,Fabian2012,SanchezAlmeida2014}. Among various physical properties of ionized gas, the kinematics traced by emission lines contains dynamic information that cannot be unveiled by the morphology of narrow-band images (e.g. $\nii$+$\ha$). 

In recent decades, with the application of integral field spectrographs (IFS), galactic astronomy is entering a new era. From spatially resolved spectra, a large number of two-dimensional maps regarding the properties of ionized gas are derived, including ionized state \citep{Sarzi2010,Belfiore2016}, gas-phase metallicity \citep{Sanchez2014} etc. Particularly, the velocity map of emission lines, describing the two-dimensional distribution of the line-of-sight velocity, is a projection of the internal motion of ionized gas, where its morphology provides an intuitive way to characterize the gas kinematics \citep{Sarzi2006,Glazebrook2013}. 

For an ideal rotational supported disk galaxy, the regular rotational motion leads to a symmetric and spider-like pattern in the velocity map (See the right-most panel of Figure \ref{fig:vasym_case} for an example). However, the velocity maps of real galaxies are not perfectly symmetric but exhibit more or less asymmetric features. Such asymmetry features  indicate that non-rotational motion is a non-negligible component of galaxy kinematics. With such a scenario, the asymmetric degree of velocity map, dubbed kinematic asymmetry, is a good indicator of the degree of non-rotational motion. 

Many studies have shown that some types of galaxies, such as merging galaxies \citep{Shapiro2008,Simons2019,Feng2020}, barred galaxies \citep{Wong2004,Sellwood2010,Holmes2015} and AGN host galaxies \citep{Liu2013,Slater2019}, can display asymmetric velocity maps. This phenomenon indicates those types of galaxies contain more intense non-rotational motion, which is linked with tidal disturbance, bar-driven gas inflow, and AGN-driven gas outflow. However, non-rotational motion within other kinds of galaxies is poorly understood due to a lack of observational studies. 

Recent IFS galaxy surveys, such as CALIFA \citep{CALIFA} and SAMI \citep{SAMI}, have built a large galaxy sample spanning a wide range of properties, allowing for a good census of velocity maps to all types of galaxies. Based on the SAMI survey, \citet{Bloom2017,Bloom2018} performs a systematical study for the kinematic asymmetry of $\ha$ velocity map and found the dependency of kinematic asymmetry on stellar mass, star formation, galaxy interaction, and gas content. While due to the limitation of the small sample size, their research doesn't analyze merging galaxies and non-merging galaxies separately, so the properties of kinematic asymmetry for non-merging galaxies are still unknown. Besides, it is also not well addressed how observational effects, such as signal-to-noise ratio of emission lines and spatial resolution, influence kinematic asymmetry. It is necessary to perform a more detailed study based on a larger galaxy sample. 

As the largest IFS galaxy survey, SDSS-IV MaNGA survey \citep{Bundy2015} contains about $10,000$ galaxies, which provides an excellent opportunity to perform further statistical studies of kinematic asymmetry. In the first paper of this work series, we construct a catalog of kinematic asymmetry of $\ha$ velocity map for all MaNGA galaxies and present the general properties of kinematic asymmetry. This paper is organized as follows. In Section \ref{sec:data}, we first introduce the galaxy sample and measurement of kinematic asymmetry. Then we investigate the kinematic asymmetry of three types of `special' galaxies in Section \ref{sec:cases}. For other types of galaxies, their properties of kinematic asymmetry are presented in Section \ref{sec:normal}. In Section \ref{sec:dis}, we discuss the influence of the observational effect on kinematic asymmetry. In the end, we give a brief summary in Section \ref{sec:sum}. Through our paper, we adopt cosmological model with $h=0.7$, $\Omega_0=0.3$ and $\Omega=0.7$. 

\section{Data}\label{sec:data}

\subsection{MaNGA Survey}

The MaNGA galaxy survey is one of the three key projects of SDSS-IV \citep{Blanton2017}. It aims to obtain spatially resolved spectroscopic data with an IFS instrument mounted by the 2.5-m telescope at Apache Point Observatory \citep{Smee2013,Drory2015,Gunn2006}. Since the targeted galaxies covered a wide range of stellar mass and colors \citep{Yan2016b,Wake2017} with a median redshift of $z = 0.03$, it provides a representative sample of galaxies in the local universe. 

We adopt the data of MaNGA Product Launch 11 (MPL-11) containing $10590$ unique galaxy observations, which is identical to the public data release in SDSS DR17 \citep{SDSSDR17}. The data are reduced by the MaNGA Data Reduction Pipeline (DRP, \citealt{Law2015,Yan2016a}) and Data Analysis Pipeline (DAP; \citealt{Westfall2019,Belfiore2019}), and provide spatially resolved distribution of various derived properties, such as the line-of-sight velocity and integrated flux of emission lines. In this work, we adopt the maps of derived properties from the DAP pipeline for analysis.

\subsection{Definition of Kinematic Asymmetry}\label{sec:kin}

\begin{figure*}
    \centering
    \includegraphics[width=\textwidth]{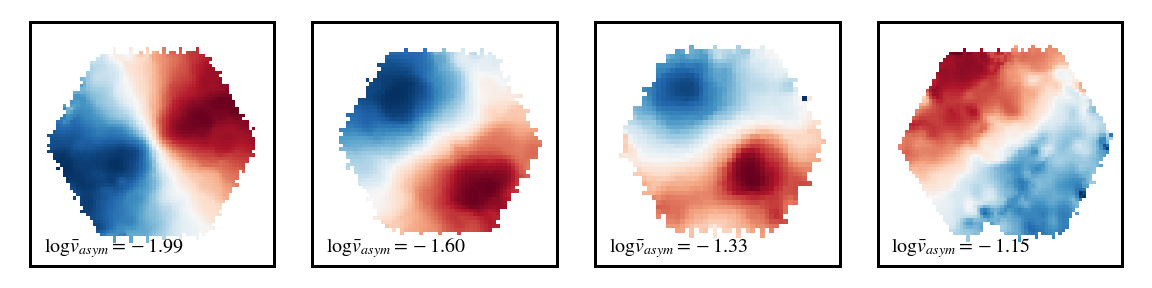}
    \caption{Four cases of $\ha$ velocity map, the kinematic asymmetry of which $\logv$ are labelled in the bottom left of each panel. }
    \label{fig:vasym_case}
\end{figure*}

We use the \texttt{kinemetry} package \citep{kinemetry} to measure the asymmetric degree of the morphology of the velocity map \footnote{\url{http://www.davor.krajnovic.org/software/}}. This package divides a velocity map into a sequence of concentric elliptical annuluses. Each annulus is determined by three parameters, position of the galaxy center, position angle, and ellipticity. Within the given annulus, \texttt{kinemetry} fit the velocity value by the Fourier series, which is expressed as
\begin{equation}\label{eq:vff}
    V(a, \psi) = A_0(a) + \sum^N_{n=1} [A_n(a)\sin(n\psi) + B_n(a)\cos(n\psi)] ,
\end{equation}
In this equation, $\psi$ is the azimuthal angle, $a$ is the semi-major axis length of ellipse annulus, and the $A_0$ represents the systematic velocity. For convenience, we could further define the amplitude, 
\begin{equation}
    k_n(a)=\sqrt{A_n^2 + B_n^2},
\end{equation}
and phase coefficients,
\begin{equation}
    \phi_n=\arctan \frac{A_n}{B_n}.
\end{equation}
Then, the Equation \ref{eq:vff} can be rephrased in more compact expression, 
\begin{equation}
    V(a,\psi)=A_0(a)+\sum^N_{n=1}k_n(a)\cos[n(\psi-\phi_n(a))]
\end{equation}
In this equation, the first-order coefficient $k_1$ describes the symmetric pattern in the velocity map, which is typically linked with the regular rotational motion. While high-order coefficients $k_n\ (n=2,3,4,5)$ describe the asymmetric pattern of the velocity map, indicating the contribution of non-rotational motion. Then, we define kinematic asymmetry as follows to quantity asymmetry degree of velocity map \citep{kinemetry,Shapiro2008}. 
\begin{equation}\label{eq:vasym}
    \vasym=\frac{k_{2}+k_{3}+k_{4}+k_{5}}{4k_{1}}\,.
\end{equation}
Here, the $\vasym$ is a function of semi-major axis length $a$, which represents the kinematic asymmetry at a given radius. For the entire galaxy, we adopt the average value of kinematic asymmetry within 1 $\re$ (all radii have the same weights) as the characteristic value, expressed as $\vasymbar$.

\subsection{Measurement of Kinematic Asymmetry}

We take the velocity maps and velocity error maps of $\ha$ emission line from MaNGA DAP \footnote{Here we adopt the version of \texttt{HYB10-MILESHC-MASTARHC2}, where the parameters of emission lines are obtained by single Gaussian fitting. } to measure kinematic asymmetry. We first remove $134$ galaxies where the galaxy center is not consistent with the IFS bundle center and $1417$ galaxies whose effective radius is larger than the MaNGA bundle size. For each velocity map, we only consider the `good spaxels', where the SNR of $\ha$ is greater than $5$ and the mask of data quality for $\ha$ emission line equals zero. Galaxies whose `good spaxels' fraction within 1$\re$ is less than $90\%$ are removed to avoid fitting patchy velocity maps. The effective radius $\re$ is adopted as the \texttt{NSA\_ELPETRO\_TH50\_R} in NSA catalog \footnote{\url{http://www.nsatlas.org/}}. Finally, $5353$ galaxies are kept in our sample, denoted as Sample A. Most of them are star-forming disk galaxies. 

During fitting, the center of concentric elliptical rings is fixed as the central spaxel of the MaNGA velocity map. The width of the elliptical annulus is set as $1\arcsec$, and the inner radius of the first ring is set as $1.5\arcsec$. The spaxel size of MaNGA maps is $0.5\arcsec$. We only fit those elliptical annuluses where the `good spaxels' fraction is larger than $75\%$ (the parameter \texttt{cover} in \texttt{kinemetry} is set as $0.75$). The position angle (PA) and ellipticity ($q$) of concentric elliptical rings are determined based on the velocity map. Specifically, we first fit the velocity map by setting both PA and $q$ as free parameters and calculating the median values of PA and $q$ which vary with radius. Then we fit the velocity maps again by fixing PA and $q$ is the median value of the first round fitting. The resulted $\vasymbar$ values of our sample galaxies are ranging from $0.008$ to $4.955$ ($-2.10$ to $0.70$ in log scale). For convenience, we adopt logarithm value $\logv$ in the following discussion.

In Figure \ref{fig:vasym_case}, we show the $\ha$ velocity maps and corresponding kinematic asymmetry $\logv$ of four cases. In each velocity map, only `good spaxels' which are used for estimating kinematic asymmetry are shown. From left to right, we can see that the morphologies of velocity maps begin to deviate gradually to form a symmetric pattern. Correspondingly, the values of kinematic asymmetry also increase from left to right. 

\subsection{Uncertainty of Kinematic Asymmetry}\label{sec:vasymerr}

\begin{figure}
    \centering
    \includegraphics[width=\columnwidth]{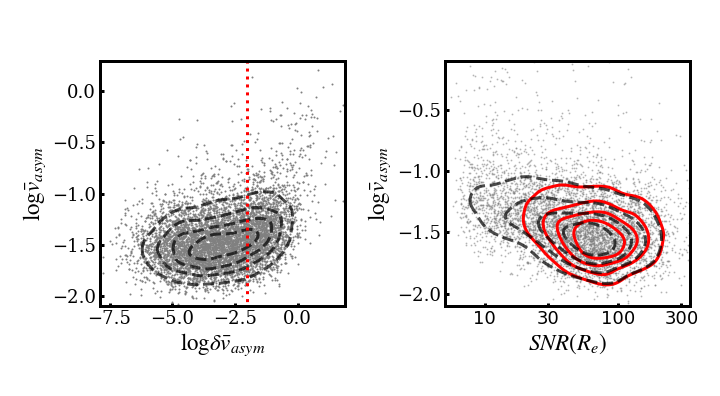}
    \caption{\textit{Left}: The relationship between kinematic asymmetry $\logv$ and its relative uncertainty $\dlogv$. Gray dots are all $5353$ galaxies in Sample A, and the black dashed contours show their number density distribution. The red dotted line represents $\dlogv=-2$. 
    \textit{Right}: The relationship between kinematic asymmetry $\logv$ and signal-to-noise ratio of $\ha$ flux at effective radius $\snr$. The black dashed contours represent the number density distribution of Sample A, while red solid contours represent galaxies satisfying $\dlogv<-2$. }
    \label{fig:vasym_err}
\end{figure}

Before investigating the statistical properties of kinematic asymmetry, we need to discuss its uncertainty. When fitting the velocity maps, \texttt{kinemetry} provides the error of all fitting parameters in Equation \ref{eq:vff}. Then, the characteristic error of kinematic asymmetry within $1\re$ ($\verr$) can be estimated by the propagation of uncertainties. Here, we adopt the relative error $\delta \vasymbar =  \verr/\vasymbar$ to quantify the uncertainty of kinematic asymmetry. Since this value is typically smaller than $1$, for convenience we use the logarithm value $\dlogv$ for discussion.

The left panel of Figure \ref{fig:vasym_err} shows the value of kinematic asymmetry as a function of its relative error. The gray dots represent all $5353$ galaxies in Sample A, the black dashed contours represent the number density distribution, and the red dotted line represents the location of $\dlogv = -2$ where the uncertainty of kinematic asymmetry equals $1\%$ of kinematic asymmetry. We notice that, for galaxies on the left side of the red line, the distribution of $\logv$ is approximately independent of the relative error. While for galaxies on the right side, the kinematic asymmetry tends to have a larger value as its relative uncertainty increases.

There are two reasons for the correlation between kinematic asymmetry and its relative uncertainty. First, galaxies with $\dlogv > -2$ indeed have more asymmetric velocity maps. Their highly asymmetric features can not be fully described by five order Fourier series, so their $\dlogv$ increases. Second, this phenomenon is caused by the signal-to-noise ratio (SNR) of $\ha$ flux. Generally, the poorer SNRs of emission lines can lead the larger values in velocity error maps, which will further increase the uncertainty of kinematic asymmetry. On the other hand, the lower SNRs can induce the larger fluctuation of the best-fit velocities \citep{Belfiore2019}, which produce additional sub-structures in velocity maps. When fitting velocity maps, those additional sub-structures will increase the weights of higher-order components in Equation \ref{eq:vff}, then increase the value of $\logv$.

We make a simple simulation to illustrate this point. For the galaxy $8313$-$9102$, the second galaxy in Figure \ref{fig:vasym_case}, we artificially add random fluctuations with a level of its median velocity error ($\sim 2.5\kms$) to the velocity maps. Then we measure the kinematic asymmetry and find that the value of kinematic asymmetry is $0.03$dex larger than before. If the added fluctuation is as large as $10\kms$ which corresponds to the expected error when $\ha$ emission-line has $\text{SNR}\sim 10$, the increase of $\logv$ will reach $0.09$dex. Such an effect is much more significant for the galaxies in which the intensity of the rotational components is weaker (e.g. face-on galaxies). For example, the kinematic asymmetry of galaxy $8728$-$12701$, the fourth galaxy in Figure \ref{fig:vasym_case}, will be increased by $0.17$dex after adding a random fluctuation of $10\kms$.

The right panel of Figure \ref{fig:vasym_err} illustrates that poor SNR of $\ha$ flux is the main reason for large $\dlogv$. In this plot, the x-axis is the median value of $\ha$ SNR of all spaxels around $\re$, denoted as $\snr$, which represents the characteristic SNR of $\ha$ flux for the whole galaxy. Similar to the left panel, we use gray dots to represent all $5353$ galaxies and use black dashed contours to display their number density distribution. The red solid contours show the number density distribution of galaxies with $\dlogv<-2$. The most distinct difference between the two contours is that the black ones cover more galaxies with low $\snr$ (e.g. $\snr<20$) than the red ones. That means galaxies with low $\snr$ indeed have a larger relative error of kinematic asymmetry.  Moreover, the galaxies with $\snr < 20$ have a larger $\logv$ value than others, consistent with our expectation. 

To reduce the influence of $\ha$ SNRs on the investigation of kinematic asymmetry, we only considered galaxies with $\dlogv<-2$ (denoted as Sample B) in the followings, including $3905$ galaxies. We have also tested that slightly changing the threshold of $\dlogv$ or using $\snr$ as criteria does not affect any of our conclusions. 

\subsection{Distribution of Kinematic Asymmetry}

\begin{figure*}
    \centering
    \includegraphics[width=\textwidth]{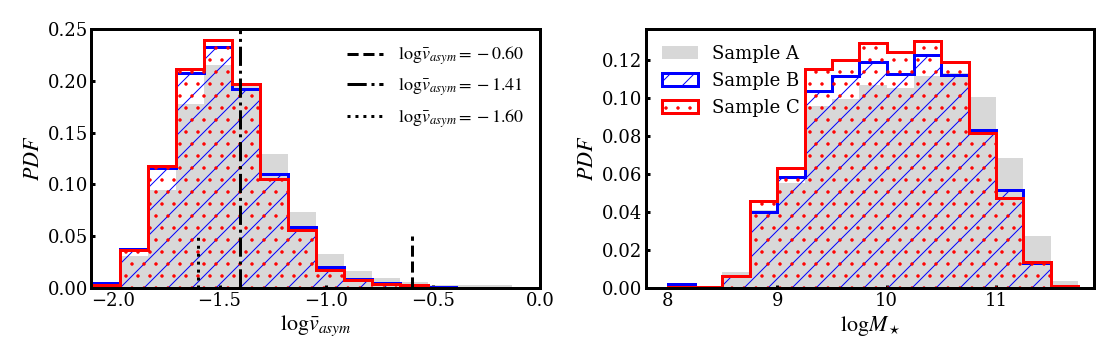}
    \caption{\textit{Left:} The distribution of kinematic asymmetry for Sample A, B and C. $\logv=-1.41$ (dotted-dashed line) is the $66$th percentiles of $\logv$ distribution, which is adopted as the threshold of high asymmetry (HA) galaxy. $\logv=-0.60$ (dashed line) and $\logv=-1.60$ (dotted line) indicate when the ratio between non-rotating component and rotating component equals $1$ and $0.1$ respectively. \textit{Right:} The distribution of stellar mass for Sample A, B and C. }
    \label{fig:vasym_pdf}
\end{figure*}

In the left panel of Figure \ref{fig:vasym_pdf}, we present the kinematic asymmetry distribution of Sample B with a blue hatched histogram. Generally speaking, the distribution shows a continuous Gaussian-like profile with a small tail on the side of large $\logv$. The median $\logv$ of Sample B is $-1.50$. The $16$th and $84$th percentile of $\logv$ distribution are $-1.71$ and $-1.26$ respectively. 

According to the definition of kinematic asymmetry in Equation \ref{eq:vasym}, if the intensity of non-rotating component ($k2+k3+k4+k5$) in velocity maps equals to the rotating component ($k1$), the value of kinematic asymmetry is $\logv = -0.60$ (dashed line in Figure \ref{fig:vasym_pdf}). If the intensity of non-rotating component is $10\%$ of the rotating component, $\logv$ equals to $-1.60$ (dotted line). For Sample B, only $15$ galaxies ($0.4\%$) satisfy $\logv>-0.6$ where the non-rotational motions dominate the internal kinematics. In the other $99.6\%$ galaxies the rotational motions play the dominant role. Among them, $1275$ galaxies ($37.8\%$) have $\logv < -1.6$ and $2615$ galaxies ($67.2\%$) have $-1.6 <\logv <-0.6$. Such a result means that the rotational motion plays the dominant role whereas the contribution of non-rotational motions cannot be ignored for most star-forming disk galaxies in the local universe. 

From this plot, we can see that the contribution of the non-rotational component in the velocity map varies about two orders of magnitude. The correlation between the non-rotational motion and other physical properties of galaxies is indeed the main target of our study. To simplify the investigation, we follow the method of \citet{Feng2020} and define galaxies with $\logv>-1.41$, which is the $66$th percentiles of $\logv$ distribution and is displayed by a black doted-dashed line, as high kinematic asymmetry (HA) galaxies. Next, we explore these HA galaxies from two aspects. We first examine the  kinematic asymmetry distribution (fraction of HA galaxies) for sub-samples of galaxies that are already known to be special in some respects (Section \ref{sec:cases}). We then further explore which type of galaxies are these HA galaxies in general (Section \ref{sec:normal}). 

\section{Kinematic Asymmetry of `Special' Galaxies} \label{sec:cases}

As mentioned in Section \ref{sec:intro}, merging galaxies, barred galaxies, and AGN host galaxies, were reported to have asymmetric velocity maps. For convenience, we use the term `special' galaxy to represent them. However, besides merging galaxies, whether barred galaxies and AGN host galaxies hold more peculiar velocity maps is still in debate \citep{Barrera-Ballesteros2014,Bloom2017}. Therefore, we are going to calculate the fraction of HA galaxy for each type of `special' galaxy. If the fraction is significantly higher than the general galaxy sample, it means that this type of `special' galaxy is indeed easier to produce intense non-rotational motion. 

Previous studies about asymmetric morphology of velocity maps focus almost entirely on `special' galaxies. However, it is far from known whether `special' galaxies dominate the HA population in the local universe. To answer this question, we also need to calculate the fraction of `special' galaxies in the HA population, which can tell us whether there are other types of galaxies being able to show highly asymmetric velocity maps.  

We need morphological classification information for selecting `special' galaxies. Here, we adopt the result of \citet{Dominguez2018}, which provides the morphological classification of almost $680,000$ SDSS galaxies using  the deep learning technique. This catalog was trained on the morphologies provided by Galaxy Zoo \citep{Willett2013} which are also available for the majority of MaNGA sample galaxies. Among all $3905$ galaxies in Sample B, $3221$ galaxies have the morphological classification, which is denoted as Sample C. For Sample C, there are $1036$ galaxies with $\logv>-1.41$, and the HA fraction is $32.3 \pm 1.0 \%$, consistent with Sample B.

We show the distribution of kinematic asymmetry and stellar mass of Sample A and C in Figure \ref{fig:vasym_pdf} for comparison, where the stellar mass is taken from MPA-JHU catalog \footnote{\url{https://www.sdss.org/dr12/spectro/galaxy_mpajhu/}} \citep{Kauffmann2003,Salim2007}. Generally, Sample B and C have almost the same distributions of both kinematic asymmetry and stellar mass, confirming that there is no selection effect for the morphological classification catalog. But the distributions of Sample A bias to larger stellar mass and higher kinematic asymmetry, implying that the most of removed galaxies in Sample B are massive red galaxies lacking strong emission lines. Here we state that all of the analyses in the following are based on Sample C. 

\begin{table}[t]
\renewcommand{\arraystretch}{1.1}
\caption{Summary of Kinematic Asymmetry Sample}
\label{tab:KinSample}
\centering
\begin{threeparttable}
    \begin{tabular}{lrrr}
        \toprule
        Sample             & Number & HA Number & HA Fraction       \\ \midrule
        Sample A \tnote{1} & $5353$ &    $2201$ & $41.1 \pm 0.9 \%$ \\
        Sample B \tnote{2} & $3905$ &    $1304$ & $33.4 \pm 0.9 \%$ \\
        Sample C \tnote{3} & $3221$ &    $1036$ & $32.1 \pm 1.0 \%$ \\
        \bottomrule
    \end{tabular}
    \begin{tablenotes}
        \item[1] Galaxies having $\logv$ measurement in full MaNGA sample.
        \item[2] Galaxies satisfying $\dlogv<-2$ in Sample A.
        \item[3] Galaxies having morphology classification of \cite{Dominguez2018} in Sample B.
    \end{tablenotes}
\end{threeparttable}
\end{table}

\subsection{Merging Galaxies}\label{sec:merge}

We construct the merging galaxy sample by two approaches, selecting galaxies with disturbed morphology and selecting galaxies with close neighbors. First, we select galaxies with disturbed morphology according to parameter $\pmerge$ in \citet{Dominguez2018}, which represents the probabilities of galaxies having merger signatures, such as tidal tails. In this step, we obtain $525$ galaxies with $\pmerge>0.5$, named as the disturbed sample. Second, we select galaxies with close neighbors and named the pair sample. We explore neighbor galaxies around each Sample C galaxies in the SDSS main galaxy sample. To minimize the spectral incompleteness due to fiber collisions, we supplement redshift from other spectral surveys, such as the LAMOST spectra survey \citep{Feng2019}. In this step, we identify $289$ galaxies having bright neighbors ($r<17.77$) within projected separation of $\pdis<50\kpc$ and line-of-sight velocity difference of $|\dv|<500\kms$. We combined two samples to obtain the final merging galaxy sample containing $728$ galaxies.  

We find that only $86$ sources in the disturbed sample are included in the pair sample. There are several reasons for this phenomenon. First, the spectral completeness within $55$arcsec (corresponding to $34\kpc$ at $z=0.03$) of the SDSS main galaxy sample is still significantly lower than unit \citep{Shen2016}, which leads to many close pairs missed in the paired sample. Because the morphology of such kinds of galaxies is likely to be disturbed by companions, they could be included in the disturbed sample. Second, the disturbed sample includes many merging galaxies at coalescence phases, which usually have no close neighbors and can not satisfy the criteria of the pair sample. Third, for merging galaxies at the apocenter phase, their projected separation to the closest neighbor may be larger than $50\kpc$, so that they will be excluded by paired sample \citep{Feng2020}. However, if the interaction is strong enough, those galaxies can be identified by their disturbed morphology. Fourth, some sources with weak tidal structure or mildly morphological distortion can not be identified by the morphological approach but can be included in the pair sample. Obviously, these two samples are complementary to each other. 

\begin{figure}[t]
    \centering
    \includegraphics[width=\columnwidth]{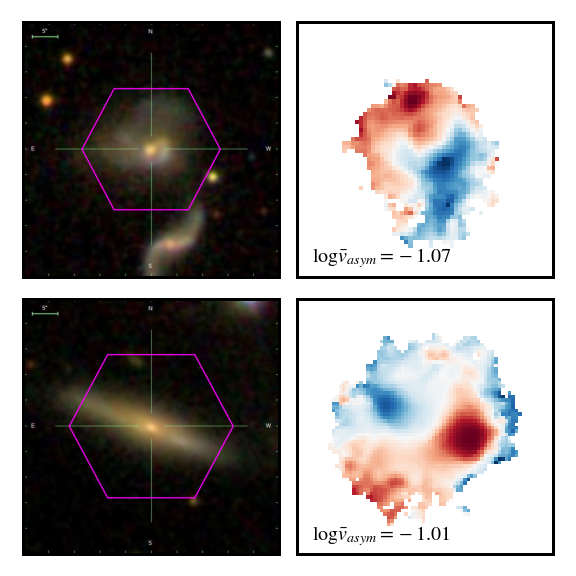}
    \caption{Two cases of merging galaxies, 10839-9102 (top), 9894-12703 (bottom). \textit{Left}: $gri$ band image from SDSS. \textit{Right}: $\ha$ velocity map, only spaxels with $\ha$ SNR greater than 5 are shown. The value of kinematic asymmetry is labelled in the bottom left corner.}
    \label{fig:case_merger}
\end{figure}

\begin{figure}[t]
    \centering
    \includegraphics[width=\columnwidth]{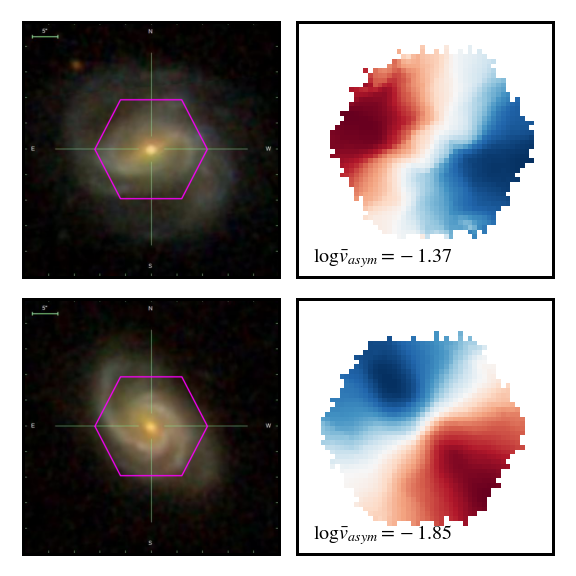}
    \caption{The cases of barred galaxy 9879-6102 (top) and unbarred disk galaxy 8317-6102 (bottom).}
    \label{fig:case_bar}
\end{figure}

Among $728$ merging galaxies, we find $276$ galaxies are HA galaxies, which makes the HA fraction of merging galaxies reach $37.9 \pm 2.3 \%$, higher than Sample C. We note that this value is the lower limit of HA fraction because the merging galaxy sample contains many non-interacting galaxies due to the projection effect \citep{Feng2020}. For comparison, the HA fraction of non-merging galaxies (other galaxies excluded by the merging sample) is $30.5 \pm 1.1 \%$ ($760 / 2493$). Obviously, merging galaxies are indeed easier to produce intense non-rotational motion.

Numerical simulation suggests that asymmetric velocity maps of merging galaxies are caused by the galaxy-galaxy interaction \citep{Mihos1996,Lotz2008,Hung2016}. When two galaxies reach the pericenter, the tidal force between them produces an additional acceleration on both gas and stars. Such an additional acceleration first disturbs the morphology of both gas velocity maps and stellar velocity maps by producing additional non-rotational velocity, and later the additional non-rotational velocity of stellar component disturbs photometric morphology through the re-distribution of stars. Because the ionized gas is further influenced by the shock process during galaxy collision, the morphology of the $\ha$ velocity map is more sensitive to galaxy interactions than photometric morphology.  Galaxies with disturbed morphology must have higher kinematic asymmetry.  

The cases in Figure \ref{fig:case_merger} can illustrate this scenario. In this figure, the left panels show the $gri$ band image from SDSS. The purple hexagon represents the observation area of MaNGA. The right panels show the $\ha$ velocity maps, where only `good spaxels' are shown. The kinematic asymmetry values are labeled in the bottom left corner of the right panels. The top galaxy 10839-9102 is a typically paired galaxy with a close neighbor on the bottom right. Its photometric morphology exhibits a prominent tidal feature, indicating this galaxy just experienced interaction. As expected, its velocity map also displays a highly disturbing pattern. And the value of kinematic asymmetry $\logv=-1.07$ implies that the non-rotational motion induced by galaxy interaction is comparable with rotational motion. The bottom galaxy 9894-12703 is also a paired galaxy, which displays a regular morphology but has a companion at $27\kpc$ away. The asymmetric velocity map implies that galaxy interaction has already disturbed gas kinematics, however, the photometric morphology has not changed yet. 

\subsection{Barred Galaxies}\label{sec:bar}

We construct barred galaxy sample as follows. First, we select $1664$ disk galaxies from Sample C with criteria of $P_{\text{disk}}>0.5$ and $\logm>9.5$. The parameter $P_{\text{disk}}$ is the probability that a galaxy is classified as a disk galaxy in the catalog of \citet{Dominguez2018}. The stellar mass threshold is adopted because low mass galaxies can not form stellar disks. Next, we remove edge-on disk galaxies whose $P_{\text{edge-on}}$ (the probability of being classified as a edge-on galaxy) value is larger than $0.5$, and $1138$ galaxies remains. Then, we adopt $P_{\text{bar}}>0.5$ (the probability of having a bar-like structure) as the thresholds to select barred galaxies. The morphology catalog provides two type of $P_{\text{bar}}$, $P_{\text{bar,GZ2}}$ which is based on the result of Galaxy Zoo 2 \citep{Masters2011} and $P_{\text{bar,Nair}}$ which is based on the result of \citet{Nair2010}. We have tested that the final result is independent of the choice of $P_{\text{bar}}$. Here, we adopt the $P_{\text{bar,GZ2}}$ in the following discussion. Finally, we obtain $231$ barred galaxies. 

In this barred galaxy sample, $115$ galaxies are belonging to HA. That makes the HA fraction of barred galaxies reach $49.8 \pm 4.6\%$. For comparison, we also investigate face-on normal disk galaxies, satisfying $P_{\text{disk}}>0.5$, $P_{\text{edge-on}}<0.5$ and $\logm<9.5$. Among $1138$ galaxies, there are $273$ HA galaxies, which make the HA fraction reach $24.0 \pm 1.5\%$. The result indicates that barred galaxies also hold more asymmetric velocity maps and contribute to a large number of high asymmetry populations in the local universe. 

The correlation between bar structure and kinematic asymmetry can be explained by the gas inflow along the bars. Many works have demonstrated that the bar structure can drive the gas within the disk flowing into the central region. Such a radial flow is bi-asymmetric and can be expressed by the second-order term of harmonic expansion \citep{Wong2004,Sellwood2010}. According to the definition of kinematic asymmetry in Equation \ref{eq:vasym}, the increasing of the second-order term of harmonic expansion must lead to the larger value of kinematic asymmetry. Moreover, the asymmetric features caused by gas inflow are spatially correlated with the bar. 

Figure \ref{fig:case_bar} shows two cases to illustrate the influence of bar structure. The bottom galaxy 8317-6102 is a typical spiral galaxy, the velocity map of which shows a symmetric spider-like pattern. The value of kinematic asymmetry $\logv = -1.85$ indicates the non-rotating component is almost negligible, excellently agreeing with the rotating disk model. In contrast, the velocity map of the barred galaxy 9879-6102 (bottom) also shows a roughly symmetric pattern but contains an additional twist feature in the inner region leading to the kinematic asymmetry $\logv=-1.37$ above the HA threshold. In particular, the twist feature is spatially correlated with barred structure in the photometric image. This phenomenon reinforces the scenario of bar-driven gas inflow.

\subsection{AGN Host Galaxies}\label{sec:AGN}

We use the integral flux of $\ha$, $\hb$, $\oiii$ and $\sii$ emission lines within central $2.5$ arcsec, which is taken from \texttt{DAPALL} catalog, to estimate the locations in BPT diagram. With the criteria of \citet{Kewley2006}, we identify $136$ AGNs, $199$ LINERs and $2885$ star-forming galaxies from Sample C.

\begin{figure}[t]
    \centering
    \includegraphics[width=\columnwidth]{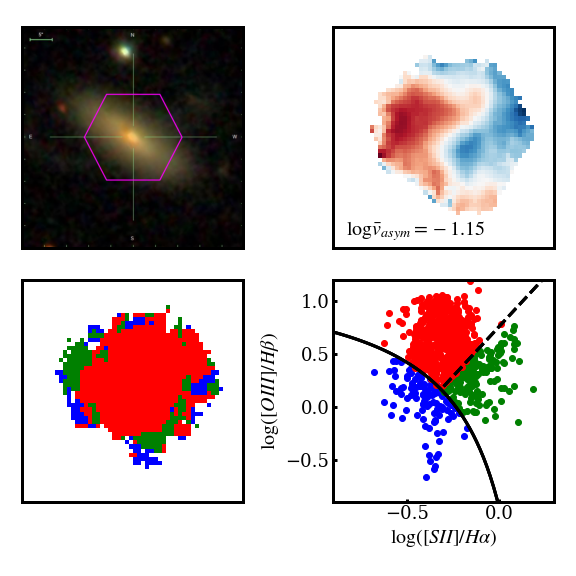}
    \caption{The case of AGN host galaxy 10492-6103. \textit{Top left}: SDSS \text{gri} band image; \textit{Top right}: $\ha$ velocity map; \textit{Bottom left}: maps of BPT classification; \textit{Bottom right}: BPT diagram for all spaxels. }
    \label{fig:case_agn}
\end{figure}

Among $136$ AGN host galaxies, there are $74$ galaxies having $\logv>-1.41$, making the HA fraction reach $54.4 \pm 6.3 \%$. For LINERs, we only find $67$ HA galaxies, corresponding to an HA fraction of $33.7 \pm 4.1$. For star-forming galaxies, the HA fraction is $31.0 \pm 1.0 \%$ ($894 / 2885$). This result confirmed that the AGN host galaxies indeed have higher kinematic asymmetry on average, while LINERs only show a marginal excess on HA fraction than star-forming galaxies. 

Higher kinematic asymmetry of AGN host galaxies can be induced by the AGN-driven outflow. Many works have demonstrated that AGN feedback can trigger a biconical outflow \citep{Das2005,Crenshaw2010}. If the direction of gas outflow is perpendicular to stellar disk \citep{Shen2010}, it will produce a symmetric velocity map in a high inclination view angle. Then the combination of rotational motion and outflow motion will result in a twist feature in velocity map \citep{Venturi2018}. For LINERs, there are two reasons to interpret their normal HA fraction. First, the activities of SMBH in LINERs are not strong enough to trigger intense gas outflow. Second, a large number of LINERs may originate from evolved stars instead of central AGN \citep{Sarzi2010,Belfiore2016}. 

\begin{table*}
\center
\renewcommand{\arraystretch}{1.0}
\caption{Summary of overlapping of `special' galaxy categories}
\label{tab:HA_Special}
\begin{threeparttable}
    \begin{tabular}{lrrrrrr}
        \toprule
        Sample  & $N_\text{Total}$ & $N_\text{Merger}$ & $N_\text{Bar}$ & $N_\text{AGN}$ &
        $N_\text{LINER}$ & $N_\text{Pure}$\tnote{1} \\ \midrule
        Merger  & $728$ & $-$  & $85$ & $38$ & $55$ & $564$ \\
        Bar     & $231$ & $85$ & $-$  & $18$ & $13$ & $129$ \\
        AGN     & $136$ & $38$ & $18$ & $-$  & $ 0$ & $ 88$ \\
        LINER   & $199$ & $55$ & $13$ & $0$  & $ -$ & $137$ \\
        \bottomrule
    \end{tabular}
    \begin{tablenotes}
        \item[1] Galaxy number after removing overlapping categories.
    \end{tablenotes}
\end{threeparttable}
\end{table*}

The velocity map of 10492-6103 shown in Figure \ref{fig:case_agn} is consistent with this scenario. According to the photometric image (top left panel), this galaxy doesn't have any peculiar features, such as bars or tidal tails. However, its velocity map (top right panel) is highly disturbed and the kinematic asymmetry has reached $\logv = -1.15$, indicating very intense non-rotational motion. Particularly, the velocity map in the co-planar region roughly follows the rotating disk model, while it is begin to twist in the extra-planar region. Meanwhile, we notice that the spaxels classified as AGN region have also extended to the extra-planar region, indicating that gas outflow perpendicular to the stellar disk is a reasonable explanation. 

\subsection{The Overlap Effect of `Special' Galaxy Categories}

In the above sections, we studied the HA fraction of `special' galaxies and confirmed that all of them indeed hold stronger non-rotational motion. It should be mentioned that these categories of `special' galaxies are not mutually exclusive. For example, many AGN host galaxies are also merging galaxies, whose asymmetric velocity maps might be caused by galaxy interaction instead of AGN-driven gas outflow. Such a phenomenon may bias our final results. Then in this section, we discuss the overlap effect of `special' galaxy categories.

First, in the merging galaxy sample, there are $85$ barred galaxies, $38$ AGNs, and $55$ LINERs. After removing the above galaxies, $564$ pure merging galaxies remain. Among them, $200$ galaxies are HA galaxies, indicating the HA fraction is $35.5 \pm 2.5\%$. Second, among $262$ barred galaxies, there are $85$ merging galaxies, $18$ AGNs, and $13$ LINERs. The number of pure barred galaxies is $129$, among which $64$ galaxies are HA galaxies. Then the HA fraction is $49.6 \pm 6.2\%$. Third, for the AGN host galaxy sample, there are $38$ merging galaxies and $18$ barred galaxies. Among $88$ pure AGN galaxies, there are $49$ HA galaxies, indicating the HA fraction is $53.4 \pm 7.8\%$. Forth, for $199$ LINERs, there are $55$ merging galaxies and $13$ barred galaxies. Among $137$ pure LINERs, there $49$ HA galaxies resulting in a HA fraction of $35.8 \pm 5.1\%$. For convenience, we list overlapping numbers in Table \ref{tab:HA_Special}.

Compared with results in previous sections, we find that the HA fraction of barred galaxies, AGN host galaxies, and LINERs is unchanged within $68\%$ confidence intervals after removing overlapping. Even though the HA fraction of the merging sample after removing overlapping indeed decreases, which is still larger than the whole Sample C and also non-merging samples. Therefore, our conclusion that `special' galaxies have a larger HA fraction is unchanged. The result of the HA fraction for `special' galaxies are listed in Table \ref{tab:fha}.

\subsection{Fraction of `Special' Galaxies in High Asymmetry Population}

In this section, we investigate how many HA populations in Sample C are `special' galaxies. Among $3221$ galaxies, there are $1036$ galaxies satisfying $\logv > -1.41$. According to previous sections, the total number of HA galaxies in the merging sample, barred sample, AGN, and LINER sample are $276$, $115$, $74$, and $67$ respectively. These four types of galaxies contribute $26.6 \pm 1.6\%$, $11.1 \pm 1.0\%$, $7.1 \pm 0.8\%$, and $6.5 \pm 0.8\%$ HA population to Sample C. Considering the overlap of four categories, the total number of HA galaxy is $442$. That means `special' galaxies only contribute $42.6 \pm 2.0 \%$ HA populations in Sample C. 

We use `regular' galaxies to denote those galaxies excluded by the `special' galaxy sample in Sample C. Among $2122$ `regular' galaxies, there are $594$ galaxies satisfying the HA threshold. The HA fraction of `regular' galaxies is only $27.9 \pm 1.1 \%$, much lower than `special' galaxies. But they contribute $57.3 \pm 2.4\%$ HA populations in Sample C. In the local universe, more than half of galaxies with highly asymmetric velocity maps are quite `regular'. 

\begin{table}[t]
\renewcommand{\arraystretch}{1.0}
\caption{Summary of High Kinematic Asymmetry Galaxy}
\label{tab:fha}
\centering
\begin{threeparttable}
    \begin{tabular}{lrrrr}
        \toprule
        Sample  & $N_\text{Total}$ & $N_\text{HA}$ & $f_\text{HA}$ & $C_\text{HA}$ \tnote{1} \\ \midrule
        Merger\tnote{2}  & $728$ & $276$ & $37.9 \pm 2.3\%$ & $26.6 \pm 1.6\%$ \\
                & $564$ & $200$ & $35.5 \pm 2.5\%$ & $19.3 \pm 1.4\%$ \\
        Bar     & $231$ & $115$ & $49.8 \pm 4.5\%$ & $11.1 \pm 1.0\%$ \\
                & $129$ & $64$  & $49.6 \pm 6.2\%$ & $6.2 \pm 0.8\%$ \\
        AGN     & $136$ & $74$  & $54.4 \pm 6.3\%$ & $7.1 \pm 0.8\%$ \\
                & $88$  & $47$  & $53.4 \pm 7.8\%$ & $4.5 \pm 0.7\%$ \\
        LINER   & $199$ & $67$  & $33.7 \pm 4.1\%$ & $6.5 \pm 0.8\%$ \\
                & $137$ & $49$  & $35.8 \pm 5.1\%$ & $4.7 \pm 0.7\%$ \\
        \midrule
        Special & $1099$ & $442$ & $40.2 \pm 1.9\%$ & $42.6 \pm 2.0\%$ \\
        Regular & $2122$ & $549$ & $27.9 \pm 1.1\%$ & $57.3 \pm 2.4\%$ \\
        \bottomrule
    \end{tabular}
    \begin{tablenotes}
        \item[1] Percentage of HA galaxies of each type to Sample C. 
        \item[2] For `special' galaxies, the first row is the result of galaxies satisfying criteria of given category; the second row is the result of galaxies after removing overlapping categories. 
    \end{tablenotes}
\end{threeparttable}
\end{table}

\section{Kinematic Asymmetry of `Regular' Galaxies}\label{sec:normal}

There are many previous works focusing the kinematic asymmetry of `special' galaxies \citep[e.g. ][]{Shapiro2008,Holmes2015,Bloom2017,Bloom2018,Feng2020}. In contrast, the asymmetric velocity maps in `regular' galaxies have not been systematically studied. And the associated non-rotational motion is far from unknown. 

In this section, we will explore the correlation between kinematic asymmetry and other general physical properties, such as stellar mass, photometric morphology, and star formation. Based on those dependencies, we may shed light on the physical origin of non-rotational motion within `regular' galaxies.

\subsection{Stellar Mass}\label{sec:mass}

First of all, we investigate the dependency on stellar mass which is the most fundamental quantity for understanding galaxies. We first make a Spearman's rank correlation test for the relationship between kinematic asymmetry and stellar mass. The result ($\rho=-0.391$) indicates the kinematic asymmetry is anti-correlated with stellar mass, consistent with previous studies \citep{Bloom2017}.

\begin{figure*}[t]
    \centering
    \includegraphics[width=\textwidth]{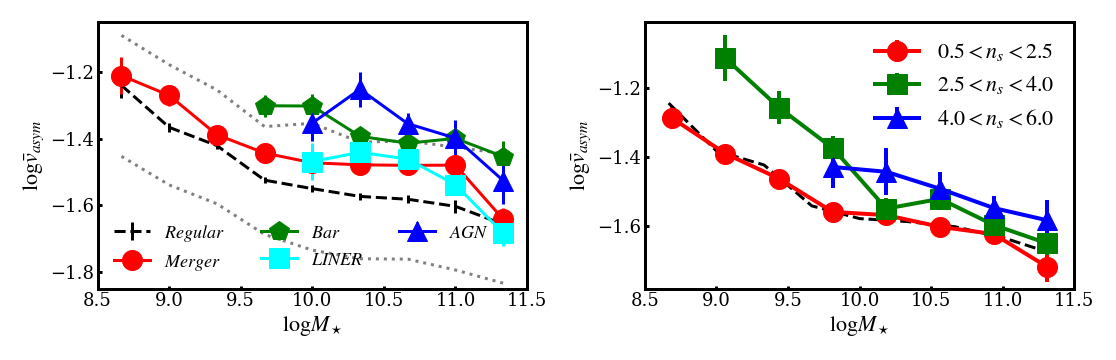}
    \caption{\textit{Left}: The kinematic asymmetry as a function of stellar mass for `regular' galaxies. Black dashed line shows the median $\logv$ values, and two black dotted lines show the $16\%$ and $84\%$ percentiles of $\logv$ distribution. The color coded symbols represent the result of `special' galaxies for comparison. \textit{Right}: Dependency of $\logm-\logv$ relation on Sersic index ($n_s$). The black dashed line is the same with left panel. }
    \label{fig:vasym_mass}
\end{figure*}

The detailed relationship between kinematic asymmetry and stellar mass is shown in the left panel of Figure \ref{fig:vasym_mass}. The black dashed line represents the median $\logv$ for given $\logm$ intervals, and the error bars show the error of median $\logv$, which is estimated by $\sigma/\sqrt{N}$. Here $\sigma$ is the standard deviation and $N$ is the galaxy number within each stellar mass bin. 

Generally, the galaxies with lower stellar mass have larger kinematic asymmetry \citep{Bloom2017}. Furthermore, the $\logm - \logv$ relation exhibits a transition mass around $\logm = 9.7$. When $\logm > 9.7$, the kinematic asymmetry only shows a very weak correlation with stellar mass. From $\logm \sim 9.7$ to $\logm \sim 11.5$, the $\logv$ only varies $0.1$dex. When $\logm < 9.7$ the kinematic asymmetry increases rapidly as $\logm$ decreasing. At $\logm \sim 8.5$, the $\logv$ reaches $-1.20$ which is $0.3$dex higher than the value at $\logm \sim 9.7$. 

Besides `regular' galaxies, we also show the $\logm-\logv$ relation of `special' galaxies in the left panel of Figure \ref{fig:vasym_mass}, including merging galaxies (red circles), barred galaxies (green pentagons), LINERs (cyan squares), and AGN host galaxies (blue triangles). Except for the largest $\logm$ bin, the median $\logv$ of `special' galaxies is higher than `regular' galaxies at a given $\logm$ interval. Because the majority of `special' galaxies are in the high mass end, the difference in the HA fraction is not as significant as the result of the $\logm-\logv$ relation. In particular, LINERs actually exhibit higher $\logv$ at given $\logm$. Because `regular' galaxies comprise many low mass galaxies which usually have larger $\logv$, the HA fraction of LINER is nearly identical to `regular' galaxies. 

Above results imply that the stellar mass is one of the key factors governing the gas kinematics. However, this is insufficient to fully determine the cause of kinematic asymmetry. As shown in the left panel of Figure \ref{fig:vasym_mass}, the scatter of $\logm-\logv$ relation is quite large ($\sim 0.2$dex), where the two dotted lines represent the $16\%$ to $84\%$ percentiles of $\logv$ distribution. It means there should be additional factors to further regulate the asymmetry of velocity maps. In the following sections, we will further examine the roles of other physical parameters, including photometric morphology, HI gas content, star formation, and environment. 

\subsection{Photometric Morphology}

\begin{figure*}[t]
    \centering
    \includegraphics[width=\textwidth]{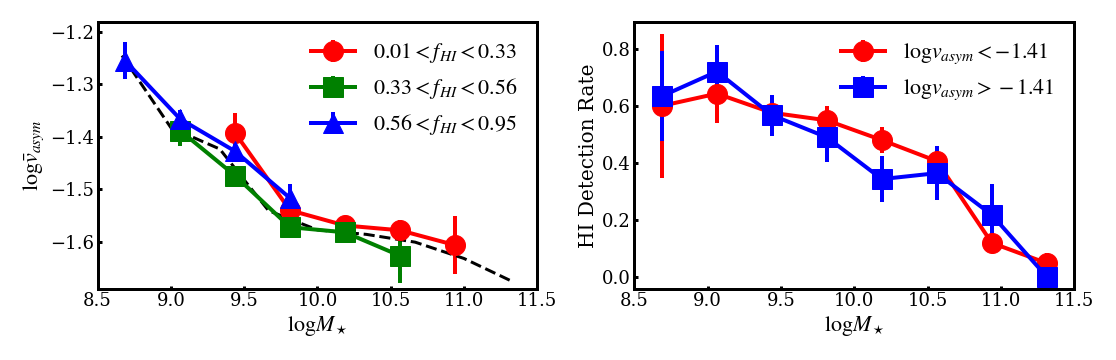}
    \caption{\textit{Left}: Dependency of $\logm-\logv$ relation on the $\hi$ fraction ($f_{\hi}$). The color coded lines represent galaxies with different $f_{\hi}$ intervals. \textit{Right}: Relationship between $\hi$ detection rate, stellar mass and kinematic asymmetry, where red circles represent galaxies with low kinematic asymmetry ($\logv<-1.41$) and blue squares represent galaxies with high kinematic asymmetry ($\logv>-1.41$). }
    \label{fig:vasym_ns}
\end{figure*}

We use Sersic index ($n_s$) to represent galaxy morphology which is taken from \texttt{SERSIC\_N} in NSA catalog, and then study the relationship between $n_s$ and $\logv$. The Spearman's rank correlation is $\rho=0.089$, indicating that there is no correlation between Sersic index and kinematic asymmetry. Because the Sersic index is anti-correlated with stellar mass, the intrinsic relationship between Sersic index and kinematic asymmetry may counteract the $\logm-\logv$ relation. To better clarify the influence of Sersic index, we need to analyze the dependency of the $\logm-\logv$ relation on Sersic index. 

We separate all `regular' galaxies into three sub-samples according to $n_s$, including $0.5<n_s<2.5$, $2.5<n_s<4.0$ and $4.0<n_s<6.0$, corresponding to disk galaxy ($1791$ galaxies), lenticular galaxy ($199$ galaxies) and elliptical galaxy ($117$ galaxies). The right panel of Figure \ref{fig:vasym_mass} shows the dependency of the $\logm-\logv$ relation on $n_s$. In this figure, data points containing fewer than $10$ galaxies are ignored. The black dashed line shows the $\logm-\logv$ relation of whole `regular' galaxies in the left panel for comparison. 

We first notice that the $\logm-\logv$ relation of $n_s<2.5$ galaxies (red circles) is nearly the same with the global trend of `regular' galaxies, because most of `regular' galaxies ($84.4\%$) are $n_s<2.5$ galaxies. While for galaxies with $n_s>2.5$, their $\logm-\logv$ relations are quite different to the black dashed line, indicating photometric morphology is indeed correlated with kinematic asymmetry. The galaxies with $4.0<n_s<6.0$ exhibit a significant excess in $\logv$ ($\sim 0.12$dex) than $n_s<2.5$ from $\logm \sim 9.5$ to $\logm \sim 11.5$. At $\logm>10.0$, those elliptical galaxies have the highest kinematic asymmetry. The galaxies with $2.5<n_s<4.0$ only show a marginal enhancement on kinematic asymmetry at $\logm>10.0$. But at $\logm<10.0$, the kinematic asymmetry of galaxies with $2.5<n_s<4.0$ is significantly higher than galaxies with $n_s<2.5$. And the difference becomes larger as the stellar mass decreasing. At $\logm = 9.0$, the median $\logv$ of galaxies with $2.5 < n_s < 4.0$ is $0.2$dex higher than galaxies with $n_s < 2.5$. 

We compare the $\snr$ of three sub-samples to verify whether the above result is biased by the signal-to-noise ratio of $\ha$ emission lines. We find that the median $\snr$ of three sub-samples at given stellar mass interval are roughly consistent, except for galaxies with $\logm > 10.8$, where the median $\snr$ of $n_s>2.5$ is lower than $50$ but larger than $30$ and the $\snr$ of $n_s<2.5$ is larger than $50$. According to right panel of Figure \ref{fig:vasym_err}, the correlation between $\logv$ and $\snr$ becomes much weak at $\snr>30$. The influence of weak emission lines on kinematic asymmetry is very limited for our result. Therefore, we may conclude galaxies with $n_s>2.5$ indeed have higher kinematic asymmetry.

\subsection{$\hi$ content}

In this section, we investigate the dependency on cold gas content, which was reported to be correlated with kinematic asymmetry \citep{Bloom2017}. We take the $\hi$ mass from $\hi$-MaNGA DR3, a 21cm follow-up program for the MaNGA survey\footnote{\url{https://www.sdss.org/dr17/manga/hi-manga/}} \citep{Master2019,Stark2021}. Through the observation of the Green Bank Telescope and ALFALFA survey \citep{Haynes2018}, $\hi$-MaNGA comprises a total of $6632$ $\hi$ observations for MaNGA galaxies.

Among $2122$ `regular' galaxies, $1030$ galaxies have $\hi$ detection. For each galaxy, the $\hi$ gas fraction is defined as $f_{\hi}=M_{\hi}/(M_{\hi}+\mass)$. The result of Spearman's rank correlation ($\rho=0.332$) indicates that $f_{\hi}$ is mildly correlated with kinematic asymmetry. However, such a correlation may originate from the $\logm-\logv$ relation because low mass galaxies are usually gas-rich and have larger kinematic asymmetry. Next, we investigate the relationship between HI mass fraction and kinematic asymmetry by fixing the stellar mass range. 

We separate all `regular' galaxies with $\hi$ detection into three equal-sized sub-samples according to $f_{\hi}$, and show their $\logm-\logv$ relations in the left panel of Figure \ref{fig:vasym_ns}. In this plot, red circles, green squares and blue triangles represent galaxies with $0.01<f_{\hi}<0.33$, $0.33<f_{\hi}<0.56$ and $0.56<f_{\hi}<0.95$ respectively. From this plot, we can see that the $\logv$ values are almost independent of gas content at any given stellar mass interval. There is no relationship between $\hi$ gas fraction and kinematic asymmetry. In the right panel, we show the relationship between $\hi$ detection rate, stellar mass, and kinematic asymmetry. In this plot, the $\hi$ detection rate is entirely determined by the stellar mass, independent of kinematic asymmetry. Based on the above two plots, we conclude that the content of cold gas does not correlate with the kinematics of ionized gas. 

\subsection{Star Formation}

\begin{figure*}
    \centering
    \includegraphics[width=\textwidth]{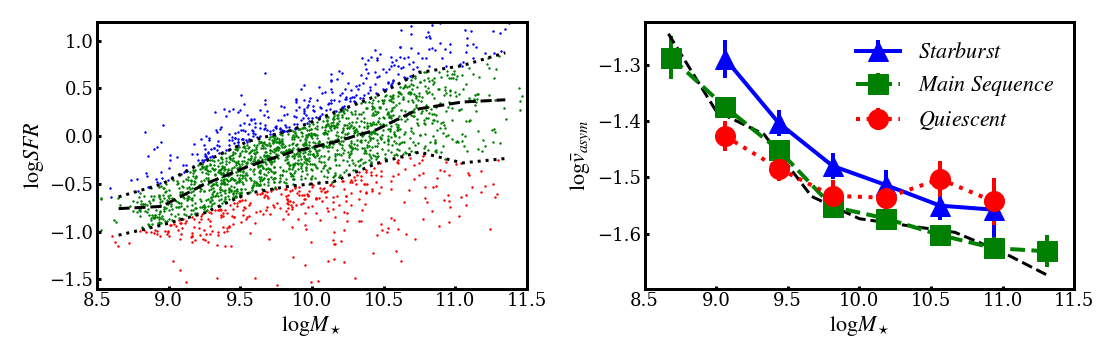}
    \caption{\textit{Left}: $\logm$ - $\log \text{SFR}$ relation of all `regular' galaxies. The black dashed line represent the median value, and the black dotted lines represent the $16$th and $84$th percentiles. \textit{Right}: The dependency of $\logm-\logv$ relation on star formation activities. Red triangles, green squares and blue circles represent starburst galaxy, main-sequence galaxy and quiescent galaxy respectively. }
    \label{fig:vasym_sfr}
\end{figure*}

In this section, we study the role of star formation. The total star formation rate (SFR) of each `regular' galaxy is taken from MPA-JHU catalog \citep{Brinchmann2004}. We perform a Spearman's rank correlation test for the relationship between specific star formation rate (sSFR, $\log \text{sSFR} = \logsfr - \logm$) and kinematic asymmetry. The result ($\rho=0.303$) implies star formation activity is also an important factor in the morphology of velocity maps. 

The left panel of Figure \ref{fig:vasym_sfr} shows the SFR as a function of stellar mass for all `regular' galaxies. The black dashed lines represent the median SFR within given $\logm$ intervals, and two dotted lines represent the $16$th and $84$th percentiles. According to this relation, we separate `regular' galaxies into three sub-samples: ones above the upper dotted line are starburst galaxies (blue), ones between two dotted lines are main-sequence galaxies (green), and ones below the lower dotted line are quiescent galaxies (red). Then we compare the $\logm-\logv$ relation of three sub-samples. 

The right panel of Figure \ref{fig:vasym_sfr} shows the $\logm-\logv$ relations of three sub-samples, where we can see that the $\logm-\logv$ relation indeed depends on star formation activities. The $\logv-\logm$ relation of main-sequence galaxies (green squares) is nearly consistent with the result of whole `regular' galaxies (black dashed line). For starburst galaxies (blue triangles), their median $\logv$ is about $0.1$dex higher than main-sequence galaxies throughout all stellar mass ranges. For quiescent galaxies (red circles), the properties of kinematic asymmetry are more complicated. When $\logm < 10$, the $\logv$ of quiescent galaxies is consistent with main-sequence galaxies. When $\logm > 10$, quiescent galaxies exhibit higher kinematic asymmetry than main-sequence galaxies, which is comparable to starburst galaxies. 

Because the $\snr$ of starburst galaxies is the highest among the three sub-samples, the result that starburst galaxies have higher kinematic asymmetry is not affected by the signal-to-noise ratio of $\ha$ emission lines. For quiescent galaxies, we find that the median $\snr$ is indeed lower than main-sequence galaxies, but still larger than $40$. In this $\snr$ range, the $\logv$ is nearly independent. Therefore, the $\logv$ of quiescent galaxies are also not biased by the signal-to-noise ratio of emission lines. 

We find that the dependency on star formation might be linked with the dependency on photometric morphology. When $\logm>10.0$, the fraction of early-type galaxies ($n_s>2.5$) of quiescent galaxies is $31.5\%$. For main-sequence galaxies and starburst galaxies, the fractions are $20.1\%$ and $22.8\%$. When $\logm<10.0$, the fraction of ETGs is the highest among starburst galaxies ($15.6\%$), while for the main-sequence and quiescent galaxies the fractions are $7.7\%$ and $8.6\%$, respectively. The $\logv$ excess of both starburst galaxies and quiescent galaxies could be explained by the higher fraction of early-type galaxies. 

In addition, the AGN feedback may also be the reason for the higher kinematic asymmetry of quiescent galaxies. Although we have excluded the AGN host galaxies, the gas outflow launched by earlier AGN feedback might still exist in `regular' galaxies, which is able to disturb the gas kinematics and quench the star formation simultaneously.

\subsection{Environment}

\begin{figure*}[t]
    \centering
    \includegraphics[width=\textwidth]{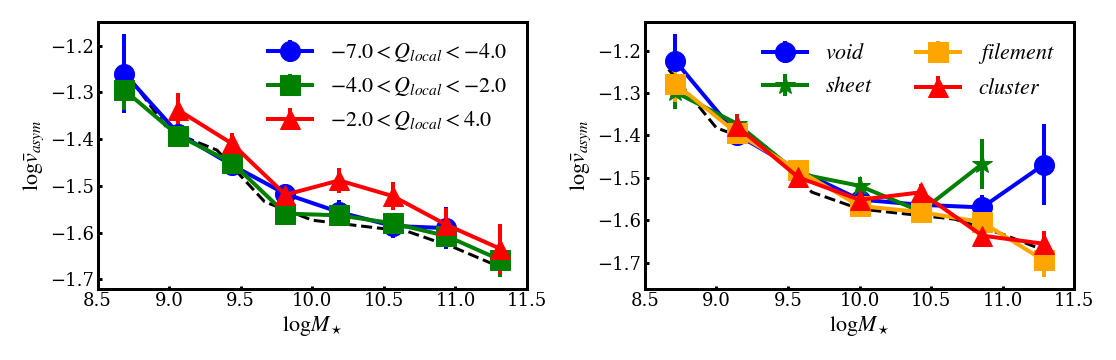}
    \caption{\textit{Left:} Dependency of $\logm-\logv$ relation on local environment ($\qlocal$): $-2.0<\qlocal<4.0$ (red triangles), $-4.0<\qlocal<-2.0$ (green squares) and $-7.0<\qlocal<-4.0$ (blue circles). 
    \textit{Right:} Dependency of $\logm-\logv$ relation on on LSS environment: void (blue circles), sheet (green stars), filament (orange squares) and cluster (red triangles). }
    \label{fig:vasym_env}
\end{figure*}

In this section, we focus on the influence of the galaxy environment on the kinematic asymmetry. At first, we examine the effect of the local environment within $1\mpc$. We use tidal strength parameter $Q$ to describe the local environment \citep{Argudo2015}, which is defined as follows, 
\begin{equation}
    \qlocal = \sum \frac{M_n}{M_p}(\frac{D_p}{\pdis})^3
\end{equation}
where $M_p$ and $D_p$ are the stellar mass and diameter of the target galaxy, $M_n$ is the stellar mass of the neighbor galaxy, and $\pdis$ is the projected separation between the target galaxy and neighbor galaxy. The neighbor galaxies include all galaxies within the intervals of $|\dv|<500\kms$ and $\pdis<1\mpc$. If the number of neighbor galaxies is larger than $5$,  $\qlocal$ only considers the closest $5$ galaxies. We take $\qlocal$ value from GEMA-VAC \footnote{\url{https://www.sdss.org/dr17/data_access/value-added-catalogs/}} sample. According to the definition of tidal strength, $\qlocal=-2$ means that the external tidal strength equals one percent of the internal bound strength. Because the `regular' galaxy sample has excluded galaxies having companions within $50\kpc$, the spectral incompleteness caused by fiber collision hardly affects the measurement of $\qlocal$. 

The Spearman's rank correlation test ($\rho=0.001$) shows that there is no correlation between $\qlocal$ and $\logv$. But when fixing the stellar mass, the $\qlocal$ exhibits a weak correlation with $\logm$. The left panel of Figure \ref{fig:vasym_env} displays the dependency of $\logm-\logv$ relation on $\qlocal$, where color coded lines represent the galaxies within $-2.0<\qlocal<4.0$ ($314$ galaxies), $-4.0<\qlocal<-2.0$ ($927$ galaxies) and $-7.0<\qlocal<-4.0$ ($417$ galaxies) respectively. In this plot, the $\logm-\logv$ relation of galaxies with $\qlocal<-2$ (blue circles and green squares) is nearly consistent with the result of whole `regular' galaxies, while the $\logm-\logv$ relation of galaxies with $\qlocal>-2$ (red triangles) lie above the black dashed line. At $\logm < 10.8$, the median $\logv$ of $\qlocal>-2$ galaxies are on average $0.05$dex larger than other two sub-samples, where the error of median $\logv$ is about $0.03$dex. But at $\logm>10.8$, there is no difference between $\qlocal>-2$ and $\qlocal<-2$.

Based on the above result, we suggest that only a high-density environment is able to impact the kinematic asymmetry. Nevertheless, we notice that both the starburst galaxies fraction and quiescent galaxies fraction of galaxies with $\qlocal>-2$ is higher than the other two sub-samples. But the fraction of early-type galaxies is roughly the same. This means the weak dependency on the local environment might be also caused by the correlation with star formation. Additionally, some `regular' galaxies are probably located in wide pairs or galaxy groups, because we only remove galaxies that have close neighbors ($\pdis<50\kpc$). Considering the effect of galaxy interaction is detectable out to $150\kpc$ \citep{Feng2020}, the higher kinematic asymmetry for galaxies with $\qlocal>-2$ might be disturbed by the neighbor galaxies within $50\kpc <\pdis<150\kpc$. 

After investigating the influence of the local environment, we next analyze the effect of the large-scale structure (LSS) environment. We adopt the definition of \citet{Hahn2007} and use the eigenvalues of tidal tensor to determine the type of structure in a cosmic web. Here, we adopt the result of GEMA-VAC catalog, which provide three eigenvalues ($T_1$, $T_2$, $T_3$) for each MaNGA galaxy by the density field reconstruction method \citep{Wang2012}. With this catalog, the LSS environment of `regular' galaxies can be identified by the eigenvalues: cluster has three positive eigenvalues; filament has two positive and one negative eigenvalue; sheet has one positive and two negative eigenvalues; void has three negative eigenvalues. 

The right panel of Figure \ref{fig:vasym_env} show the dependence of $\logm - \logv$ relation on the LSS environment. For galaxies with $\logm < 10.5$, we don't find any difference between the $\logm-\logv$ relations in different LSS environments. When $\logm > 10.5$, galaxies in the void and sheet environment show a significant excess on $\logv$ than the other two environments. We compare the other properties of galaxies with $\logm>10.5$ in four environments but don't find significant differences between them. The LSS environment may indeed impact the kinematics of ionized gas in the high mass end. 

\subsection{Physical Origin of Non-rotational Motion}

In the above sections, we have shown that the kinematic asymmetry of `regular' galaxies is correlated with stellar mass, photometric morphology, star formation activity, and environment. Next, we discuss the physical origins of kinematic asymmetry. 

First of all, the gravitational potential is the most crucial factor to regulate the kinematic status of galaxies. Generally, the galaxies with lower stellar mass typically have shallower potential, where the ionized gas is more easily disturbed and needs a longer time to settle down. As a result, the non-rotational motion will be more prominent in lower mass galaxies and then leads to a higher kinematic asymmetry \citep{Bloom2017,Bloom2018}. However, a shallow gravitational potential alone is not sufficient for producing an asymmetric velocity map, which also requires a perturbation. 

The perturbation can be induced by internal processes of galaxies. For `regular' galaxies, the stellar feedback and associated gas outflow is the most important internal process driving the perturbation of ionized gas \citep{Veilleux2005,Hopkins2012}. In the starburst region, the stellar feedback especially the explosion of supernova can expel the surrounding gas, drive the ionized gas to produce non-rotational motion, and then increases the kinematic asymmetry. \citet{Ho2016} reported that galaxies that show asymmetric velocity maps from the edge-on view, indeed exhibit enhanced star formation rate surface density. It is reasonable that the kinematic asymmetry is correlated with star formation. 

Besides those internal processes, the external processes can also perturb the gas kinematics. Among them, gas accretion is the most essential mechanism. In the current framework of galaxy formation and evolution, the gas accretion from the external environment fuels the galaxy formation and dominates their mass assembly \citep[e.g.][]{Keres2005,L'Huillier2012}. This implies that gas accretion is a very common process that can impact galaxies more frequently. On the one hand, the inflow gas from the external environment is likely to have distinct angular momentum than existing gas \citep{Davis2011,Jin2016} and then perturb the gas kinematics. On the other hand, the inflow gas may fuel the star formation or even trigger the starburst. The dependency of kinematic asymmetry on star formation could be explained by the scenario of gas accretion. Considering the dependency on star formation is linked with dependency on the Sersic index, we suspect that gas accretion might be the reason that the early-type galaxies exhibit more asymmetric velocity maps. 

In the galaxy cluster environment, the ionized gas may also be disturbed by the hot intracluster medium through ram pressure stripping. Many observations have reported that the galaxies in cluster environments show asymmetric morphologies of $\ha$ imaging \citep{Yagi2010,Boselli2018} and distinct position angles of $\ha$ kinematics to the stellar components \citep{Bryant2019}. The numerical simulations also predict that the velocity maps of ionized gas begin to exhibit asymmetric features after the galaxy enters a galaxy cluster \citep{Khim2020,Khim2021}. Therefore, the fact that galaxies in dense environments have more asymmetric velocity maps may also be interpreted by the gas stripping process.

At last, we should mention that the kinematic asymmetry of `special' galaxies is also correlated with other physical properties, such as stellar mass (see left panel of Figure \ref{fig:vasym_mass}), photometric morphology, and star formation. This indicates above the physical mechanisms we discussed in `regular' galaxies are also able to produce intense non-rotational motions in `special' galaxies. For example, the velocity maps of starburst merging galaxies are not only influenced by galaxy interaction but also regulated by gravitational potential and starburst-driven gas outflow. 

\section{Influence of Observational Effect}\label{sec:dis}

In Section \ref{sec:vasymerr}, we have illustrated that the SNR of $\ha$ flux strongly affects the measurement of kinematic asymmetry. Besides, there are other observational effects that can also bias the value of kinematic asymmetry. In this section, we discuss how those observational effects influence our final results. 

\subsection{Redshift and Spatial Resolution}

Similar to photometric morphology, the morphology of the velocity map is also strongly affected by spatial resolution. For a galaxy at $z=0.01$, the size of the resolved smallest structure in its velocity map is about $0.53\kpc$ at a given spatial resolution $2.5\arcsec$. While for a galaxy at $z=0.05$, the size of the smallest structure increases to $2.54\kpc$, five times larger than the former. If we move one galaxy to a higher redshift, the velocity map will be `smoothed', then the value of kinematic asymmetry might be changed. 

\begin{figure*}
    \centering
    \includegraphics[width=\textwidth]{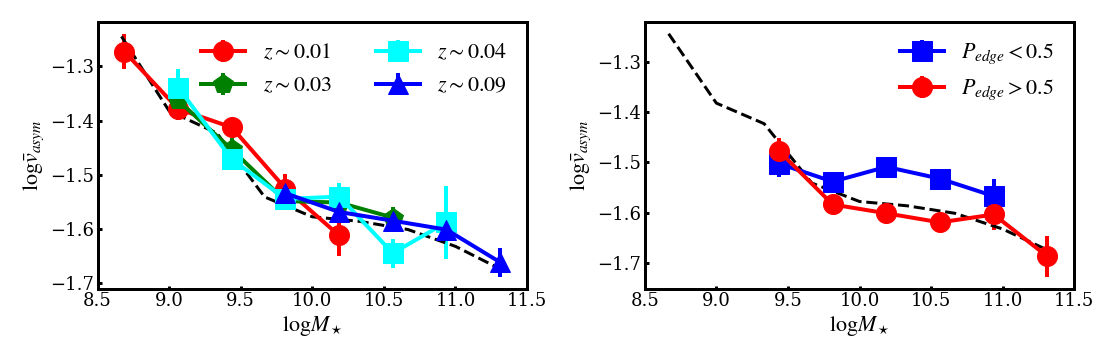}
    \caption{\textit{Left}: Dependency of $\logm-\logv$ relation on redshift. The color coded symobols represent the sub-samples defined by redshifts. \textit{Right}: Dependency of $\logm-\logv$ relation on inclination angle. The red circles represent edge-on like galaxies with $P_\text{edge}>0.5$ and $P_\text{disk}>0.5$; the blue squares represent the face-on like galaxies with $P_\text{edge}<0.5$ and $P_\text{disk}>0.5$.}
    \label{fig:vasym_z}
\end{figure*}

To evaluate the influence of this effect, we investigate the dependency of $\logm - \logv$ relation on galaxy redshift. We separate the `regular' galaxy sample into four equal-sized sub-samples according to the redshift and show their $\logm - \logv$ relation in the left panel of Figure \ref{fig:vasym_z} using red circles, green pentagons, cyan squares, and blue triangles. The median redshift of each sub-sample are $0.01$, $0.03$, $0.04$, $0.09$ respectively. We can see that the $\logm - \logv$ relations of four sub-sample are nearly identical to the whole sample, indicating that spatial resolution can not affect the value of kinematic asymmetry at least in the redshift range of the MaNGA sample. 

\subsection{Inclination}\label{sec:dis_incl}

In observation, the velocity we derived from the spectrum is the projection of intrinsic 3-dimensional velocity along the line of sight. The velocity values in the velocity map depend on the inclination angle. As a result, all fitting harmonic coefficients (from $k_1$ to $k_5$) should be strongly correlated with the inclination, which will further lead that the value of kinematic asymmetry is a function of the inclination. 

To evaluate the effect of inclination statistically, we investigate the relationship between the $\logm-\logv$ relation of `regular' galaxy and inclination. We select $1028$ disk galaxies from `regular' galaxy sample with criteria of $P_\text{disk}>0.5$ and $\logm > 9.5$. Then, we separate them into two sub-samples with a threshold of $P_\text{edge}=0.5$, including $617$ face-on like galaxies with $P_\text{edge}<0.5$ and $411$ edge-on like galaxies with $P_\text{edge}>0.5$. The right panel of Figure \ref{fig:vasym_z} displays the $\logm-\logv$ relation of two sub-samples. This plot shows that the $\logv$ value of face-on like galaxies (blue squares) at given stellar mass is on average $0.05$dex larger than the edge-on like galaxies (red circles). Galaxies with low inclination angles are more likely to exhibit more asymmetric velocity maps.

The dependency on the inclination angle may be due to the fact that some kinematic features are easier to detect at low inclinations. For example, in the top panels of Figure \ref{fig:case_bar}, the twist feature of velocity maps caused by the bar structure is obvious in the face-on view. While it would disappear if we observe the same galaxy in the edge-on view. For the `regular' galaxies, although we have excluded barred galaxies, there are still some other structures (e.g. spirals) producing the asymmetric kinematic features which are only observable at low inclinations. As a result, the kinematic asymmetry of face-on like galaxies is higher on average. 

For edge-on like galaxies, dust attenuation is able to increase the kinematic asymmetry (see appendix A for a case study). But such kind of galaxies only accounts for a small fraction, so they don't affect the statistical result.

\section{Summary}\label{sec:sum}

The velocity maps of $\ha$ emission lines observed by integral field spectrograph provide a large amount of information about the internal motion of ionized gas. Because the asymmetric patterns in velocity map morphology reflect the non-rotational motions, the asymmetry degree of velocity maps is an intuitive description of the kinematic state. 

In this paper, we make a census of asymmetric degrees for the $\ha$ velocity map based on the SDSS-IV MaNGA survey. We calculate the kinematic asymmetry through \texttt{kinemetry} package and construct a catalog of kinematic asymmetry comprising $5353$ galaxies (Sample A). We find that lower SNR of $\ha$ flux will bias kinematic asymmetry to a larger value and also increases its relative uncertainty. To avoid the influence of weak emission lines, we select $3905$ galaxies (Sample B) with low relative uncertainties of kinematic asymmetry from Sample A. From Sample B, we further select $3221$ galaxies (Sample C) with photometric morphology classification of \citet{Dominguez2018} for the following analysis. 

First, we investigate the kinematic asymmetry of `special' galaxies, including merging galaxies, barred galaxies, AGN host galaxies, and LINERs. We calculate the fraction of high kinematic asymmetry (HA, $\logv>-1.41$), which are $37.9\%$, $49.8\%$, $54.4\%$ and $33.7\%$ respectively. Considering the HA fraction of Sample C is $32.3\%$, the result indicates that those `special' galaxies are indeed easier to have stronger non-rotational motion, consistent with the scenarios about galaxy interaction, barred-driven gas inflow, and AGN-driven gas outflow. However, we find that only $43.8\%$ of HA populations in Sample C are `special' galaxies, which implies that other types of galaxies dominate HA galaxies in the local universe. 

We use `regular' galaxies to denote other types of galaxies in Sample C and study the statistical properties of kinematic asymmetry. We find that stellar mass shows an anti-correlation with kinematic asymmetry. We find a transition mass $\logm = 9.7$, above which the kinematic asymmetry only shows a very weak correlation with the stellar mass. When $\logm < 9.7$, the kinematic asymmetry increases rapidly with decreasing in stellar mass. For a given stellar mass, the kinematic asymmetry also shows a correlation with other physical properties, such as the Sersic index, star formation rate, and local density environment. While for the galaxies with HI detection, we don't find that the kinematic asymmetry is correlated with HI gas content. 

In the end, we discuss the influence of the observational effect on kinematic asymmetry. We find that inclination angle can affect the value of kinematic asymmetry, where face-on galaxies have more asymmetric velocity maps. On the other hand, we don't detect the correlation between redshift and kinematic asymmetry, indicating that the physical spatial resolution is an irrelevant factor inside the MaNGA redshift coverage.

The result that `regular' galaxies dominate the HA population in the local universe, implies that non-rotational motion of ionized gas is very common during galaxy evolution. According to the dependency of kinematic asymmetry, we suggest that gas accretion/inflow and gas outflow are able to perturb the gas kinematics. However, such a general investigation is insufficient to clarify the detailed roles of those two processes. In our next paper, we will study the links between kinematic asymmetry and chemical abundance to disentangle the effect of gas inflow and outflow. And the results will provide a new constraint on the origin of non-rotation motion, but also a new insight into gas cycling within the galaxy ecosystem.

\acknowledgments

This work is supported by National Natural Science Foundation of China (No. 12103017, 12073059, U2031139, 11933003), Natural Science Foundation of Hebei Province (No. A2021205001), Project of Hebei Provincial Department of Science and Technology (No. 226Z7604G), Postdoctoral Research Program of Hebei Province (No. B2021003017), Science Foundation of Hebei Normal University (No. L2021B08) and the National Key R$\&$D Program of China (No. 2019YFA0405501, 2017YFA0402704). We also acknowledge the science research grants from the China Manned Space Project with NO. CMS-CSST-2021-A04, CMS-CSST-2021-A05, CMS-CSST-2021-A07, CMS-CSST-2021-A08, CMS-CSST-2021-A09, CMS-CSST-2021-B04. F.T.Y. acknowledges support by the Funds for Key Programs of Shanghai Astronomical Observatory (No. E195121009) and the Natural Science Foundation of Shanghai (Project Number: 21ZR1474300). 

Funding for the Sloan Digital Sky Survey IV has been provided by the Alfred P. Sloan Foundation, the U.S. Department of Energy Office of Science, and the Participating Institutions. SDSS-IV acknowledges support and resources from the Center for High-Performance Computing at the University of Utah. The SDSS website is \url{www.sdss.org}.

SDSS-IV is managed by the Astrophysical Research Consortium for the Participating Institutions of the SDSS Collaboration including the Brazilian Participation Group, the Carnegie Institution for Science, Carnegie Mellon University, the Chilean Participation Group, the French Participation Group, Harvard-Smithsonian Center for Astrophysics, Instituto de Astrof\'isica de Canarias, The Johns Hopkins University, Kavli Institute for the Physics and Mathematics of the Universe (IPMU) / University of Tokyo, the Korean Participation Group, Lawrence Berkeley National Laboratory, Leibniz Institut f\"ur Astrophysik Potsdam (AIP),  Max-Planck-Institut f\"ur Astronomie (MPIA Heidelberg), Max-Planck-Institut f\"ur Astrophysik (MPA Garching), Max-Planck-Institut f\"ur Extraterrestrische Physik (MPE), National Astronomical Observatories of China, New Mexico State University, New York University, University of Notre Dame, Observat\'ario Nacional / MCTI, The Ohio State University, Pennsylvania State University, Shanghai Astronomical Observatory, United Kingdom Participation Group, Universidad Nacional Aut\'onoma de M\'exico, University of Arizona, University of Colorado Boulder, University of Oxford, University of Portsmouth, University of Utah, University of Virginia, University of Washington, University of Wisconsin, Vanderbilt University, and Yale University.

\newpage

\begin{figure*}[h]
    \centering
    \includegraphics[width=0.75\textwidth]{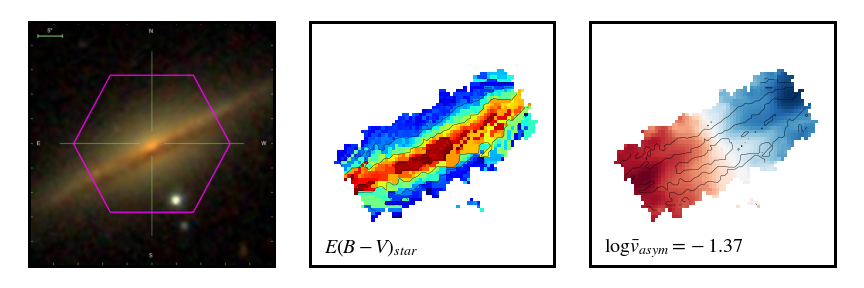}
    \caption{The case of dusty edge-on galaxy 9040-12701. The panels from left to right are SDSS image, $\ebv$ map and $\ha$ velocity map. The gray contours in velocity map represent the $\ebv$ distribution. }
    \label{fig:case_dust}
\end{figure*}

\appendix
In this appendix, we show that dust attenuation will also increase kinematic asymmetry. Figure \ref{fig:case_dust} shows one case of dusty edge-on galaxies. From the photometric image , we notices that this galaxy has an obvious dust lane. That produces a belt-like feature with large $\ebv$ in dust attenuation map taken from \textsc{PIPE3D} \citep{PIPE3Da,PIPE3Db,PIPE3Dc}. The morphology of $\ha$ velocity map is not symmetric and the velocity deviation compared to rotating disk model are spatially correlated with dust attenuation. For convenience, we use gray contours to represent the $\ebv$ distribution in the right panel. We notice that the spaxels with larger $\ebv$ exhibit more significant deviation, consistent with the prediction of \citet{Baes2003}. Compared to other edge-on galaxies, the galaxies with obvious dust lanes may have larger kinematic asymmetry. 

\bibliography{ref}{}
\bibliographystyle{aasjournal}

\end{CJK*}
\end{document}